# Chip-to-chip photonic quantum teleportation over optical fibers of 12.3km

Dongning Liu,[1] Zhanping Jin,[1] Jingyuan Liu,[1] Xiaotong Zou,[1] Xiaosong Ren,[1] Hao Li,[2] Lixing You,[2] Xue Feng,[1] Fang Liu,[1] Kaiyu Cui,[1] Yidong Huang,[1,3] and Wei Zhang,[1,3*]

[1] *Frontier Science Center for Quantum Information, Beijing National Research Center for Information Science and Technology (BNRist), Electronic Engineering Department, Tsinghua University, Beijing 100084, China.*

[2] *National Key Laboratory of Materials for Integrated Circuits, Shanghai Institute of Microsystem and Information Technology, Chinese Academy of Sciences, Shanghai 200050, China.*

[3] *Beijing Academy of Quantum Information Sciences, Beijing 100193, China.*

*\* Corresponding author. E-mail address: zwei@tsinghua.edu.cn.*

**Abstract:**

Quantum teleportation is a crucial function in quantum networks. The implementation of photonic quantum teleportation could be highly simplified by quantum photonic circuits. To extend chip-to-chip teleportation distance, more effort is needed on both chip design and system implementation. In this work, we demonstrate a chip-to-chip photonic quantum teleportation over optical fibers under the scenario of star-topology quantum network. Time-bin encoded quantum states are used to achieve a long teleportation distance. Three photonic quantum circuits are designed and fabricated on a single chip, each serving specific functions: heralded single-photon generation at the user node, entangled photon pair generation and Bell state measurement at the relay node, and projective measurement of the teleported photons at the central node. The unbalanced Mach-Zehnder interferometers (UMZI) for time-bin encoding in these quantum photonic circuits are optimized to reduce insertion losses and suppress noise photons generated on the chip. Besides, an active feedback system is employed to suppress the impact of fiber length fluctuation between the circuits, achieving a stable quantum interference for the Bell state measurement in the relay node. As the result, a photonic quantum teleportation over optical fibers of 12.3km is achieved based on these quantum photonic circuits, showing the potential of chip integration on the development of quantum networks.

1. **Introduction**

Quantum teleportation transfers quantum states through prior-distributed entanglement, providing a way to deliver quantum information without transmit its physical carrier. Since it was proposed in 1993[1], quantum teleportation has been realized in various quantum systems[2–7]. Photons, as "flying" qubits, are good carriers of quantum information to achieve long-distance quantum teleportation. The photonic quantum teleportation has been demonstrated over indoor fiber spools of 102 km[8], metropolitan fiber cables[9–13] and free space between satellite and ground[14]. Quantum teleportation is an essential function in the quantum network with a star topology[9]. In this network, a central node serves as a quantum information processor and many user nodes send their quantum information to the central node as the inputs of quantum information processing. If the user nodes are not convenient to send the photons to the central node directly, the photonic quantum teleportation in a quantum relay configuration[15,16] can be used to realize the quantum information transmission. In this configuration, the user node only needs to send the photons to a nearby relay node, which has prior-distributed entanglement resources with the central node. The quantum information can be further transmitted from the relay node to the central node via quantum teleportation. This configuration greatly reduces the transmission distance of the photons from the user nodes in a quantum network with high-quality prior-distribution of entanglement resources. Additionally, photonic devices for switching and routing could be set in the relay nodes to select a specific user node sending its quantum information, enabling the multi-user access to the central node of the quantum network.

In recent years, quantum photonic circuits developed rapidly. By integrating multiple optical components onto a single chip, compact and stable integrated quantum photonic systems can be developed to realize complicated and large-scale quantum information functions, such as complicated entanglement state generation[17–20], quantum communication[21–24], quantum computation and simulation[25–27], and so on. It can be expected that the physical implementation of photonic quantum teleportation also can be highly simplified by integration technologies of quantum photonic circuits. Previous works have demonstrated the feasibility of quantum teleportation between chips, using path-encoded quantum states on the quantum photonic chips, which are transferred to polarization-encoded quantum states in optical fibers between them. The transmission distance in these quantum teleportation experiments are about 10 meters[28,29], far from the requirement of quantum networks. More effort is needed to extend the distance of quantum teleportation between chips, including chip design and system implementation.

In this work, we demonstrate a chip-to-chip photonic quantum teleportation over optical fibers under the scenario of star-topology quantum network. To achieve a long teleportation distance over optical fibers, time-bin encoded quantum states are used. Three photonic quantum circuits are designed and fabricated on a single chip, each serving specific functions: heralded single-photon generation at the user node, entangled photon pair generation and Bell state measurement (BSM) at the relay node, and projective measurement of the teleported photons at the central node. The unbalanced Mach-Zehnder interferometers (UMZI) for time-bin encoding in these quantum photonic circuits are optimized to reduce insertion losses and suppress noise photons generated on the chip. Besides, an active feedback system is employed to suppress the impact of fiber length fluctuation between the circuits, achieving a stable quantum interference for BSM in the relay node. As the result, the teleportation distance over optical fibers is 12.3km in this work, showing the feasibility of chip-to-chip photonic quantum teleportation to be implemented in metropolitan fiber networks.

## 2. Results
### 2.1 Quantum photonic circuits for the user, relay and central nodes

A scenario of photonic quantum teleportation in a star-topology quantum network is considered in this work, which is shown in Fig. 1(a). The user node sends photons with specific quantum states to a relay node. The relay node generates photon pairs with an entangled Bell state, one photon from each pair used for joint BSMs with photons from the user node and the other photon sent to the central node. In the central node, the state of teleported photons from the relay node are transformed according to the BSM results, recovering the quantum states of the photons sent by the user node. In this work, we demonstrate it in a simplified way, in which the state of a photon sent to the central node is determined by the BSM result. The quality of the state indicates whether the quantum teleportation process is successful or not. Functions of the user node, the relay node and the center node are shown in Fig. 1(b). To characterize the states of photons sent to the central node, a projective measurement system is set at the central node to performs quantum state tomography (QST). Three silicon photonic quantum circuits are designed to realize the three nodes respectively, which are depicted in Fig. 1(c)-(e).

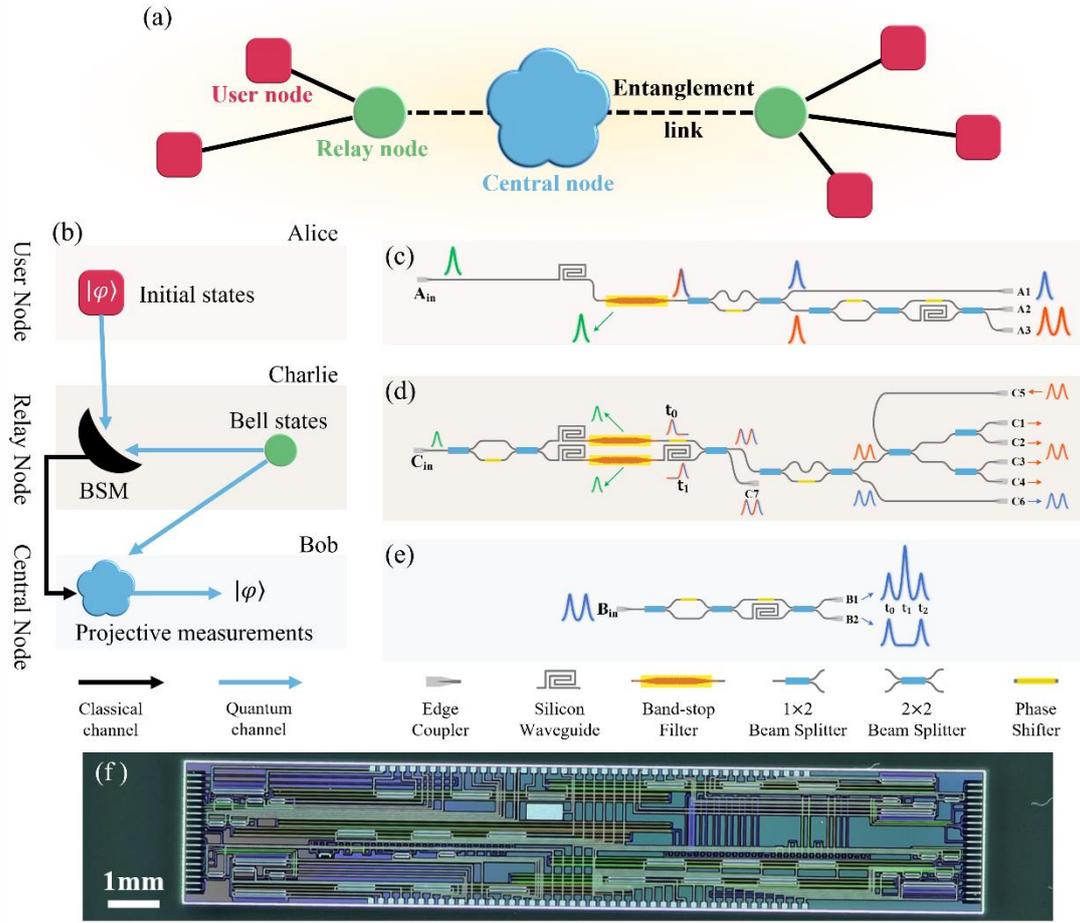

Fig 1. (a) Scenario of a star-topology quantum network. (b) Functions of the photonic quantum teleportation involving the user node, relay node and central node. (c)-(e) Designs of the silicon photonic circuits for the user, relay, and central nodes, implementing the main functions required for quantum teleportation. (f) Photograph of the fabricated silicon photonic chip, which integrated all the three quantum photonic circuits. Its size is 3 mm × 16 mm.

The quantum photonic circuit of the user node generates heralded, arbitrary time-bin encoded single-photon states, as shown in Fig. 1(c). It needs a pulsed pump light, which is coupled into the chip through the left-side edge coupler. Correlated photon pairs are generated via spontaneous four-wave mixing (SFWM) in a long single-mode silicon waveguide with a cross-section of 500 nm×220 nm and a length of 1 cm. Subsequently, a waveguide Bragg grating (WBG) band-stop filter[30–32] is used to filter out the pump light, preventing noise photons generated by the residual pump light. An unbalanced Mach-Zehnder interferometer with a free spectral range (FSR) of 1600 GHz (referred to as the 1600GHz-UMZI hereafter) separates the signal and idler photons in the generated photon pair into two paths. The signal photons output from the port A1 directly, which are used to generate heralding signals. The idler photons enter an unbalanced Mach-Zehnder interferometer with an optical path difference of 400 ps (referred to as the 400ps-UMZI hereafter), which generates the time-bin encoded single photon state $|\psi\rangle = \alpha|0\rangle + \beta e^{i\phi}|1\rangle$. When the photons pass through the short arm of the 400ps-UMZI, they are encoded into the state of earlier time-bin $|0\rangle$, while they are encoded into the state of the later time-bin $|1\rangle$ if they pass through the long arm. A thermo-optic phase shifter (TOPS) in the 400ps-UMZI adjusts the phase $\phi$ in the superposition state. It is worth noting that the length of the long arm in the 400ps-UMZI is as long as 3.2 cm. Reducing the transmission loss of the long arm and balancing the two arms are crucial for the chip-to-chip photonic teleportation experiment. Two designs are applied. First, the straight parts of the long arm are realized by rib waveguides with a rib width of 1.5 μm and a depth of 130 nm. Their propagation losses are about 0.2 dB/cm. The transition losses between the rib waveguides and single-mode strip waveguides are negligible through optimizing the transition section. Second, a variable beam-splitter (VBS) formed by a balanced MZI is used as the first beam-splitter in the 400ps-UMZI, adjusting the amplitudes of $|0\rangle$ and $|1\rangle$ in the state. In this work, this type of UMZIs with VBSs are applied in several places, which is referred to VBS-UMZI hereafter. Details of the low-loss shallow rib waveguides and the WBG band-stop filter are introduced in the supplementary information S1.

The quantum photonic circuit of the relay node generates photon pairs of time-bin encoded entangled Bell states, and performs BSM between the idler photons of the photon pairs and the received photons from the user node, and sends the signal photons to the central node, as shown in Fig. 1(d). It also needs a pulsed pump light, which is coupled into the chip through the edge coupler and injected into a VBS-UMZI. The VBS splits the pump light into the two arms. The short arm includes a single-mode silicon waveguide with a length of 1cm and a WBG band-stop filter. The pump light generates photon pairs with a biphoton state via SFWM in this waveguide, then suppressed by the WBG band-stop filter to avoid additional noise photons in the following photonic circuit. The long arm also includes a single-mode silicon waveguide with a length of 1cm and a WBG band-stop filter to generate photon pairs with a biphoton state. Additionally, the long arm has a delay waveguide of 400 ps. The VBS-UMZI has two output ports, one of them is used to superpose the biphoton states generated in the short arm and long arm coherently with a time difference of 400ps, forming the time-bin encoded entangled state $|\Phi\rangle = \frac{1}{\sqrt{2}}(|0_s 0_i\rangle + e^{i\theta}|1_s 1_i\rangle)$. Here, "0" and

"1" also denote the earlier and later time bins. "s" and "i" denote the signal and idler photons in photon pairs, respectively. A TOPS is located at the short arm, which is used to control the phase $\theta$ in the state. Specifically, the Bell state $|\Phi^{\pm}\rangle$ can be generated if $\theta$ is adjusted to 0 or $\pi$. Then, the signal and idler photons of the photon pairs with the entangled Bell state are separated by a 1600GHz-UMZI. The signal photons (blue photons in Fig. 1(d)) are output from port C6 and sent to the central node for entanglement distribution. The idler photons (red photons in Fig. 1(d)) are sent to the on-chip Bell state analyzer in the quantum photonic circuit, which is formed by cascaded beam-splitters. The photons from the user node also inject into the Bell state analyzer through port C5. For the time-bin encoded biphoton states, the Bell state analyzer based on cascaded beam-splitters can distinguish $|\Psi^{\pm}\rangle = \frac{1}{\sqrt{2}}(|0,1\rangle \pm |1,0\rangle)$ from the four Bell states. By analyzing the coincidence counts at the four output ports (C1~ C4), the results of BSM can be obtained and sent to the central node. It is worth noting that the signal and idler photons also output from the other output port of the VBS-UMZI, which is coupled out of the chip directly from the port C7 and used as a reference signal to stabilize the synchronization between the user node and the relay node.

The quantum photonic circuit of the central node characterize the state of the photons sent from the relay node when they are post-selected by the BSM results. To demonstrate the performance of photonic quantum teleportation, a VBS-UMZI for projective measurements of time-bin encoded single-photon states is designed in the circuit of the central node, as shown in Fig. 1(e). The photons from the relay node are coupled into the chip through the left-side edge coupler. By analyzing the photons output from B1 and B2 ports, projective measurements and QST is carried out, which are described in the Method section. It is worth noting that in the relay node and central node, the long arms of VBS-UMZIs are also realized by low loss shallow rib waveguides.

The three quantum photonic circuits are integrated into a silicon photonic chip with a size of 3 mm × 16 mm. The chip is fabricated by standard processes of silicon photonics compatible with complementary metal oxide semiconductor (CMOS) processes (Advanced Micro Foundry, Singapore). Figure 1(f) is the photograph of the chip. The inputs and outputs of the pump lights and generated photons are realized by the optical coupling between a fiber array and edge couplers on chip. The chip is packaged on a thermoelectric cooler (TEC) to achieve a temperature stability of ±0.002°C. Details on the insertion losses of the chip are provided in the supplementary information

S2.

## 2.2 Experimental system for photonic quantum teleportation

We use the packaged chip to perform the experiment of chip-to-chip photonic quantum teleportation over optical fibers of 12.3 km, the experimental system is shown in Fig. 2. In the user node, the quantum photonic circuit generates heralded time-bin encoded single-photon states. The wavelengths of the pump light, signal photons and idler photons are 1542.14 nm, 1538.98 nm and 1545.32 nm, respectively, corresponding to the International Telecommunication Union (ITU) channels of C44, C48, and C40. The pump light of the user node is generated by a pulsed fiber laser system with a repetition rate of 100 MHz. In the quantum photonic circuit of the user node, photon pairs are generated by SFWM, in which the signal photons are coupled out of the chip directly and used to generate the heralding signal locally. They are filtered out by an optical filter centered at the channel of C48 and detected by a superconducting nanowire single-photon detector (SNSPD, PHOTEC Inc.) with an efficiency of ~80%. The output of the SNSPD is loaded on a continuous wave (CW) laser light on the channel of C44 (Keysight N7714A) by an electro-optical intensity modulator, which is used as the optical heralding signal for the idler photons sent to the relay node. Its intensity is controlled by a variable optical attenuator to avoid noise photons generated by the optical heralded signal when it propagates in optical fibers. The idler photons of photon pairs are used to generate time-bin encoded single photon states on the chip, which shows a high fidelity over 99% in the experiment. Details of the pulsed fiber laser system and the performance of the time-bin encoded single photon state generation are introduced in the supplementary information S3 and S4. The idler photons with the time-bin encoded single-photon states and their optical heralding signals are multiplexed by a dense wavelength division multiplexer (DWDM) with a bandwidth of 200 GHz, and co-propagated over non-zero dispersion shifted fibers (G.655, Yangtze Optical Fibre and Cable) of 6.15 km to the relay node. Since the optical heralding signal also can be used to realize clock synchronization between the user node and the relay node, no additional synchronization signal is required between them.

In the relay node, the quantum photonic circuit generates entangled photon pairs with time-bin encoded Bell states using the pulsed pump light from the same pulsed fiber laser system with the user node. The idler photons of the photon pairs are detected locally for the BSM and the signal photons are sent to the central node over another spool of G.655 fibers of 6.15km. The performance of entanglement distribution between the relay node and central node is demonstrated by Franson-type interference, which shows a fringe visibility over 96% (excluding noise photons). Detailed results of the entanglement distribution are introduced in the supplementary information S5. In the relay node, the photons from the user node and their optical heralding signal passes through a variable optical delay line (ODL-1500PS, SC-Lightsource Inc.), then separated by a DWDM. The optical heralding signal is pre-amplified by an erbium-doped fiber amplifier (EDFA) and detected by a high-speed photodetector (PD). The electrical signal of the PD is recorded by the time-correlated single photon counter (TCSPC, Timetagger Ultra, Swabian Instruments) at the relay node. The photons from the user node pass through a polarization controller before they are coupled into the quantum photonic circuit of the relay node for BSM. The photons from the four output ports of the Bell state analyzer (C1 to C4) are filtered by narrow-band optical filters at the channel of C40 respectively, and then detected by SNSPDs (their detection efficiency is about 90%). Their arrival times are recorded by the TCSPC. The states projected onto the Bell states $|\Psi^+\rangle$ and $|\Psi^-\rangle$ can be distinguished by analyzing the data recorded by the TCSPC, as detailed in the Method section. Besides, the photons output from the port C7 of the relay node is filtered by a narrow-band optical filter at the channel of C48, detected by a SNSPD, and recorded by the TCSPC at the relay node. It is also worth noting that the pulsed fiber laser system provides two electrical signals for the synchronization of TCSPCs at the relay node and the central node. One is sent to the TCSPC at the relay node directly, the other is converted to an optical synchronization signal by a CW laser and electro-optical intensity modulator. The optical synchronization signal is combined with the idler photons generated in the quantum photonic circuit of the relay node by a DWDM, and then co-propagated to the central node over optical fibers of 6.15 km.

In the central node, a DWDM is used to separate the signal photons from the relay node and the optical synchronization signal. The optical synchronization signal is pre-amplified by an EDFA, detected by a PD, and recorded by the TCSPC (Timetagger Ultra, Swabian Instruments) at the central node for synchronization. The signal photons from the relay node are coupled into the quantum photonic circuit of the central node with the projective measurement system. The photons output from the ports B1 and B2. Then they are filtered by narrow-band optical filters at the channel of C48 and detected by SNSPDs. The arrival times of these single photon detection events are recorded by the TCSPC. The BSM results at the relay node and the projective measurement results at the central node are compared and analyzed, obtaining the four-fold coincidence counts to demonstrate the photonic quantum teleportation.

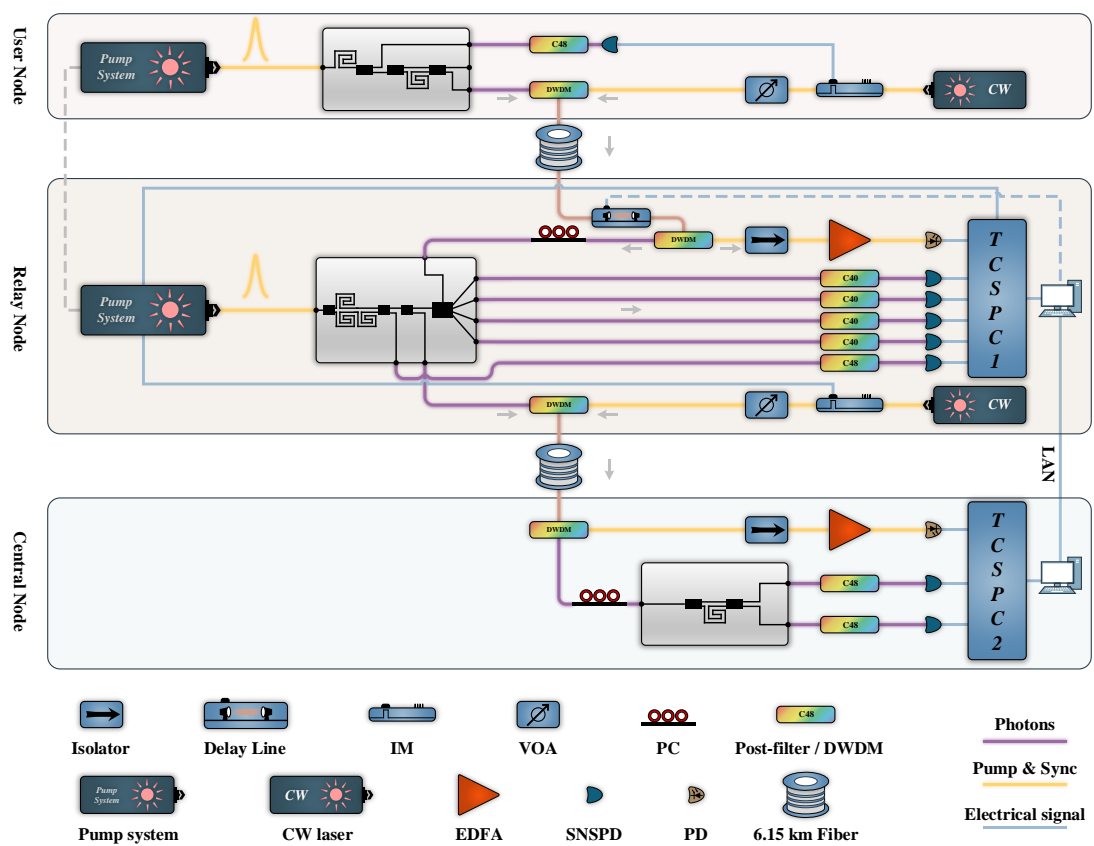

Fig 2. Experimental system of the chip-to-chip photonic quantum teleportation over optical fiber of 12.3 km. The system includes time-bin encoded single-photon state generation at the user node, time-bin encoded entangled Bell state generation and BSM at the relay node, and projective measurements at the central node. The co-propagation of quantum signals and classical synchronization signals over optical fibers ensures the stable synchronization of photon arrival times in the fiber links. IM: Intensity modulator. VOA: Variable optical attenuator. PC: Polarization controller. DWDM: Dense wavelength division multiplexer. CW: Continuous wave. EDFA: Erbium-doped fiber amplifier. SNSPD: Superconducting nanowire single-photon detector. PD: Photodetector. LAN: Local area network.

### 2.3 Stability of two-photon interference

The implementation of the photonic quantum teleportation experiment depends on the performance of BSM in the relay node. It requires that the photons from the user node and the idler photons generated locally in the relay node should be indistinguishable. However, the photons from the user node propagates along optical fibers of 6.15km, their arrival times would fluctuate due to the variation of ambient temperature. It would lead to an unstable arrival time difference, denoted as

$\Delta_{delay}$, between the single-photon wavepackets from the user node and those from the relay node in BSM. An active feedback system with a variable optical delay line is used in the experiment to measure and stabilize $\Delta_{delay}$ (see Methods for details). To show the effect of the active feedback system, the ambient temperature and $\Delta_{delay}$ are measured with and without the stabilization by the active feedback system. The results are shown in Fig. 3. Figures 3(a) and (b) show the results when the active feedback system is off. It can be seen that the fluctuation of $\Delta_{delay}$ is about 400 ps when the ambient temperature varies by 2°C. The fluctuation is significantly exceeding the coherence time of the photons from the user node and the relay node, which is about 20 ps determined by the bandwidth of the narrow-band optical filters used to select these photons. When the active feedback system is on, as shown in Fig. 3(c) and (d), $\Delta_{delay}$ no longer depends on the ambient temperature and highly suppressed with a standard deviation of ~2.2 ps over 48 hours, far smaller than the photon coherence time. It shows that the active feedback system can support stable quantum interference between the single-photon wavepackets from the user node and the relay node. It is demonstrated by the experiment of HOM interference between the photons from user node and the idler photons generated locally, using the beam splitters in the Bell state analyzer on the chip. The photons at both sides are in thermal states when they are not heralded by their signal photons. Hence, the fringe visibility of this interference has an upper limit of 33.3%. Figure 3(e) is the results of the normalized coincidence counts of the HOM interference experiment and its fitting fringe, showing a visibility (28.8%±1.2%) close to the upper limit. The stability of the two-photon interference ensures the success of the quantum teleportation experiment.

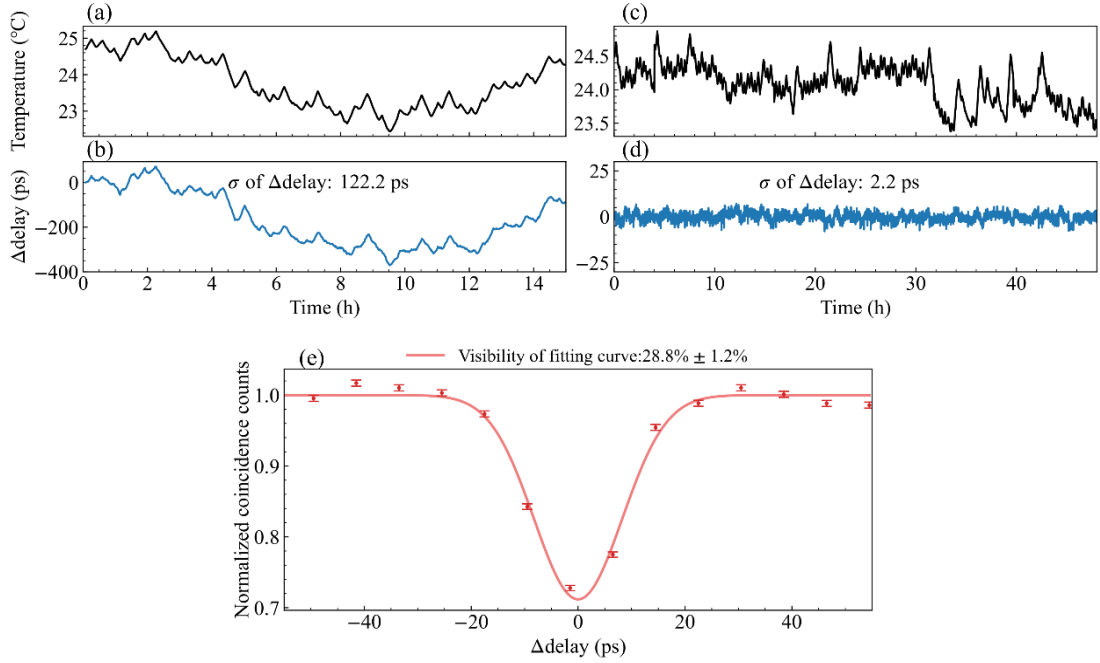

Fig 3. The performance of the active feedback system. (a) and (b) The room temperature and $\Delta_{delay}$ fluctuations without the stabilization by the feedback control. (c) and (d) Stabilized $\Delta_{delay}$ after activating the feedback system. $\sigma$ represents the standard deviation of $\Delta_{delay}$ fluctuations. (e) The normalized coincidence counts of the HOM interference experiment, with a fitted visibility of 28.8%±1.2%.

### 2.4 Results of Quantum teleportation

The performance of the chip-to-chip quantum teleportation is determined by BSM at the relay node, which is tested firstly. In the measurement, the single photon state sent by the user node is set to $|\psi\rangle = (|0\rangle + e^{i\phi}|1\rangle)/\sqrt{2}$. The phase $\phi$ could be adjusted by the TOPS at the VBS-UMZI in the quantum photonic circuit of the user node. Four-fold coincidence counts are measured under varying $\phi$ and fixed conditions of the VBS-UMZIs in the circuits of the relay node and the central node. The four ports for the four-fold coincidence measurement are the output port A1 at the user node for the heralded signal, two of the output ports for BSM at the relay node (C1~C4), and one of the two output ports for projective measurement at the central node (B1 and B2). The measurement time at each phase $\phi$ is four hours. The results when BSM projects the two photons at the relay node into the state $|\Psi^+\rangle$ and $|\Psi^-\rangle$ are plotted in Fig. 4(a) and (b), respectively. The blue and red dots are the four-fold coincidence counts when the photons at central node output from ports B1 and B2, respectively, and the blue and red lines are their fitting curves by sinusoidal functions. It can be seen that all the results show sinusoidal interference fringes with varying $\phi$, confirming that the phase information of the single photon state sent by the user node is successfully transmitted to the central node by quantum teleportation over optical fibers of 12.3km. The average visibility of these fringes is 63.3% ± 2.4%. It is worth noting that since the success probability for distinguishing $|\Psi^+\rangle$ is 50% in the Bell state analyzer based on cascaded beam splitters, the maximum coincidence counts on the interference fringes related to $|\Psi^+\rangle$ are approximately half of those related to $|\Psi^-\rangle$, which are about 46 and 90, respectively. Hence, the rate of quantum teleportation events is about 34 per hour in this chip-to-chip quantum teleportation experiment over optical fibers of 12.3 km.

Then, we prepare six input states ($|0\rangle$, $|1\rangle$, $|+\rangle$, $|-\rangle$, $|+i\rangle$, and $|-i\rangle$) at the user node and perform QST on the photons with the teleported state at the central node. The measurement time is 24 hours for each input state. When the result of BSM is $|\Psi^+\rangle$, the teleported state should be $|\psi_t\rangle = \sigma_x|\psi\rangle$; when the result of BSM is $|\Psi^-\rangle$, the teleported state should be $|\psi_t\rangle = \sigma_y|\psi\rangle$. The density matrices for the teleported states are reconstructed and their state fidelities are calculated. The experiments of photonic quantum teleportation are performed over optical fibers of 12.3km. It is also performed under a back-to-back case for comparison, in which the fiber spools are removed from the experimental system. The results are shown in Table 1. It can be seen that the fidelities of all the teleported states under fiber transmission exceed the classical limit of 2/3, with an average value of 81.2% when the result of BSM is $|\Psi^+\rangle$ and 81.0% when the result of BSM is $|\Psi^-\rangle$. It demonstrates that the quantum teleportation is successfully implemented between the circuits of the three nodes over optical fiber of 12.3km in the experimental system. The detailed density matrices are provided in supplementary information S7.

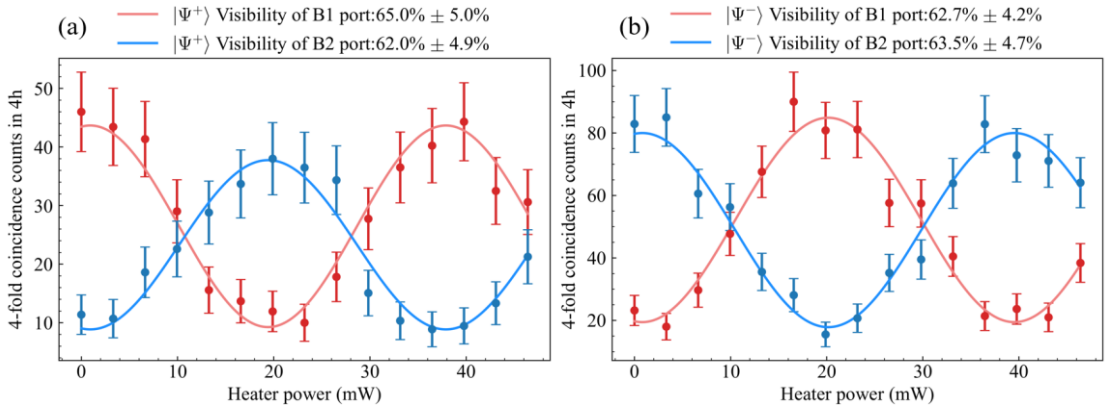

Fig 4. The experiment results of BSM when the single photon state sent by the user node is set to $|\psi\rangle = (|0\rangle + e^{i\phi}|1\rangle)/\sqrt{2}$. (a) and (b) The 4-fold coincidence counts when the results of BSM are $|\Psi^+\rangle$ and $|\Psi^-\rangle$ respectively.

Table 1. Comparison between fidelities of the quantum teleportation experiments over optical fibers of 12.3 km and those of back-to-back case.

| Input state $|\psi\rangle$ | | $|0\rangle$ | $|1\rangle$ | $|+\rangle$ | $|-\rangle$ | $|+i\rangle$ | $|-i\rangle$ | Avg. |
|---|---|---|---|---|---|---|---|---|
| Teleported state | | $|1\rangle$ | $|0\rangle$ | $|+\rangle$ | $|-\rangle$ | $|-i\rangle$ | $|+i\rangle$ | |
| BSM: $|\Psi^+\rangle \Rightarrow \sigma_x|\psi\rangle$ | 12.3 km | 84.1% | 86.4% | 81.8% | 77.0% | 78.7% | 79.3% | **81.2%** |
| | Back-to-back | 88.3% | 85.4% | 80.8% | 83.3% | 78.9% | 82.4% | **83.2%** |
| Teleported state | | $|1\rangle$ | $|0\rangle$ | $|-\rangle$ | $|+\rangle$ | $|+i\rangle$ | $|-i\rangle$ | |
| BSM: $|\Psi^-\rangle \Rightarrow \sigma_y|\psi\rangle$ | 12.3 km | 86.3% | 87.6% | 75.6% | 79.0% | 76.3% | 90.9% | **81.0%** |
| | Back-to-back | 93.1% | 88.0% | 83.6% | 78.4% | 80.1% | 83.3% | **84.4%** |

## 3. Discussion

Comparing with the results of the back-to-back case, the average fidelity of the teleported states in the experiment with optical fibers decreases by 2% when the result of BSM is $|\Psi^+\rangle$ and 3.4% when the result of BSM is $|\Psi^-\rangle$. It is mainly due to the impact of fiber dispersion on the photons from the user node when they transmit to the relay node over optical fibers of 6.15 km. The fiber dispersion would broaden the wavepacket of these photons, worsening the indistinguishability between the photons from the user node and the idler photons generated locally at the relay node when they are interfered in the Bell state analyzer in the quantum photonic circuit of the relay node. In this experiment, non-zero dispersion shifted fibers (G.655) are used, which have a dispersion parameter of 4 ps/(nm·km). The bandwidths of the optical filters for the idler photons are about 0.18 nm, leading to a temporal width of ~20 ps for the wavepacket of the photons in BSM. It can be expected that the temporal width would increase to ~25ps due to the fiber transmission. Besides, frequency chirp also would be introduced in the wavepacket due to the fiber dispersion. These effects would impact the fidelity of the teleported states. This impact could be avoided by using dispersion shifted fibers or proper dispersion compensation to extend the fiber transmission distance. It is worth noting that the fidelities of the teleported states in the back-to-back case also have spaces to be improved. They are mainly limited by the multi-pair events in the quantum light sources at the user node and the relay node, which impact the performance of BSM at the relay node. The performance could be improved if the pump lights in these quantum light sources are controlled at a lower level. Additionally, the temporal widths of the signal-photon wavepackets in the BSM could be broadened if optical filters with narrower bandwidths are used for selecting photons before detection. It is helpful to reduce the impact of residual fiber length fluctuations between the user node and the relay node on the quantum interference in the BSM, thereby increasing the fidelities of the teleported states. It can be expected the rates of successful teleportation events would be reduced if above two methods are used. Hence, quantum photonic circuits with smaller insertion losses are preferred to achieve these improvements.

## 4. Materials and methods

**4.1 Projective measurement and quantum state tomography**

In the quantum photonic circuit of the central node, a VBS-UMZI is used to perform projective measurements on the time-bin encoded single photon state. When the VBS is set to a balanced state as a 50:50 beam splitter, the input state is projected onto three consecutive time bins ($t_0 \sim t_2$) at the output ports of the VBS-UMZI, as shown in Fig.1(e). If the single photon at the central node is detected in time bins $t_0$ and $t_2$ in a four-photon coincidence event, the teleported state is projected to the measurement bases $|0\rangle$ and $|1\rangle$, respectively. The four-fold coincidence counts of these two cases are denoted by $n_0$ and $n_1$. If the single photon at the central node is detected in time bin $t_1$, which output port it is detected should be checked. The teleported state is projected to the measurement bases $|0\rangle + e^{i\alpha}|1\rangle$ or $|0\rangle + e^{i(\alpha+\pi)}|1\rangle$, determined by the photon is detected at the port B1 or B2. Here, $\alpha$ is the phase introduce by the VBS-UMZI at the central node. If $\alpha$ is set to 0, the two measurement bases are $|+\rangle$ and $|-\rangle$, respectively. The four-fold coincidence counts of these two cases are denoted by $n_+$ and $n_-$. If $\alpha$ is set to $\pi/2$, the two measurement bases are $|+i\rangle$ and $|-i\rangle$, respectively. The four-fold coincidence counts of these two cases are denoted by $n_{+i}$ and $n_{-i}$.

Based on the six projective measurement results, the density matrix of the teleported state $|\psi_t\rangle$, which is denoted by $\rho$, is reconstructed by QST. The density matrix is expressed as[9]:

$$\rho = \frac{1}{2}\sum_{i=0}^{3} S_i \sigma_i,$$
$$S_i = \langle \psi_t | \sigma_i | \psi_t \rangle,$$
$$\sigma_0 = |0\rangle\langle 0| + |1\rangle\langle 1| = \begin{pmatrix} 1 & 0 \\ 0 & 1 \end{pmatrix},$$
$$\sigma_1 = |+\rangle\langle +| - |-\rangle\langle -| = \begin{pmatrix} 0 & 1 \\ 1 & 0 \end{pmatrix}, \quad (1)$$
$$\sigma_2 = |+i\rangle\langle +i| - |-i\rangle\langle -i| = \begin{pmatrix} 0 & -i \\ i & 0 \end{pmatrix},$$
$$\sigma_3 = |0\rangle\langle 0| - |1\rangle\langle 1| = \begin{pmatrix} 1 & 0 \\ 0 & -1 \end{pmatrix},$$

In the expression, $S_i$ represents the Stokes parameters. Due to the normalization, the value of $S_0$ is 1. In experiments, the Stokes parameters are related to the projective measurement results:

$$\begin{aligned} S_0 &= \langle \psi_t | \sigma_0 | \psi_t \rangle \\ &= \langle \psi_t | 0 \rangle \langle 0 | \psi_t \rangle + \langle \psi_t | 1 \rangle \langle 1 | \psi_t \rangle \\ &= P_{|0\rangle} + P_{|1\rangle} = \frac{n_0 + n_1}{n_0 + n_1} = 1, \\ S_1 &= P_{|+\rangle} - P_{|-\rangle} = \frac{n_+ - n_-}{n_0 + n_1}, \\ S_2 &= P_{|+i\rangle} - P_{|-i\rangle} = \frac{n_{+i} - n_{-i}}{n_0 + n_1}, \\ S_3 &= P_{|0\rangle} - P_{|1\rangle} = \frac{n_0 - n_1}{n_0 + n_1}, \end{aligned} \quad (2)$$

Here, $P_{|\psi\rangle}$ represents the probability to measure the state $|\psi\rangle$.

In the experiment, two measurements are performed by setting the phase $\alpha$ of the VBS-UMZI to 0 and $\pi/2$, each for 12 hours. Thus, for a specific state sent from the user node, a measurement time of 24 hours are required to obtain the experimental data for the density matrix reconstruction. The fidelities of the teleported states are calculated by[33]:

$$F(\rho_{aim}, \rho) = [Tr(\sqrt{\sqrt{\rho_{aim}}\rho\sqrt{\rho_{aim}}})]^2, \quad (3)$$

Here, $\rho_{aim}$ is the density matrix of the teleported state assuming that the quantum teleportation process is perfect.

**4.2 Bell state measurement and HOM interference**

The quantum photonic circuit of the relay node uses a cascaded 50:50 beam splitters as the Bell state analyzer for the time-bin encoded two-photon states. The time-bin encoded Bell states, $|\Psi^+\rangle$ and $|\Psi^-\rangle$, can be distinguished by analyzing the coincidence counts of the single photon detection events at the four output ports of the Bell state analyzer (C1~C4) and two time-bins $t_0$ and $t_1$, as shown in Fig. 1(d). If the two photons are in the state $|\Psi^-\rangle$, they would be separated by the quantum interference at the first beam splitter. Hence, one photon would be detected at the port C1 or C2, the other photon would be detected at the port C3 or C4, and they would be detected at different time bins. On the other hand, if the two photons are in the state $|\Psi^+\rangle$, they would output from the same port of the first beam splitter by the quantum interference, then be separated by the second beam splitter with a possibility of 50%. In this case, the two photons would be detected at the ports C1 and C2 (or C3 and C4), respectively, and they are also detected at different time bins. The two photons also have a possibility of 50% that they would be output from the same port at the two time bins. However, since the time difference of the two time bins is too small to be discriminated by the single photon detectors, this case could not be used to distinguish $|\Psi^+\rangle$ through coincidence measurement. If the two photons in an unknown state, above coincidence measurement results could be used to indicate whether the state is projected to $|\Psi^-\rangle$ or $|\Psi^+\rangle$. In the experiment, the BSM results are post-selected by the heralded signal from the user node through the three-fold coincidence measurement. The time tags of these single photon detection events are stored and transmitted to the computer at the central node by local area network (LAN), in which they are compared with the projective measurement results of the teleported photons, completing the photonic quantum teleportation.

It can be seen that the quantum interference in the first beam splitter is crucial for the BSM. A HOM interference experiment are performed to test performance of the quantum interference. In this experiment, the photon counting events from ports C1 and C2 are combined as the events from one output port of the first beam splitter using the channel merging function of the TCSPC, while the events from ports C3 and C4 are combined as those from the other output port. The coincidence counts between the two output ports of the first beam splitter are measured in the same time bin under different time difference $\Delta_{delay}$, performing the measurement of HOM interference.

**4.3 The active feedback system for stabilization of the two-photon interference**

An active feedback system with a variable optical delay line is used in the experiment to measure and stabilize $\Delta_{delay}$, which is the arrival time difference in BSM between the single-photon wavepackets from the user node and those from the relay node. The optical heralding signal co-propagate with the time-bin encoded photons to the relay node and their time delay after the optical fiber transmission would be adjusted simultaneously by a variable optical delay line. Then, the optical heralding signal are separated by a DWDM, amplified by an EDFA, detected by a PD and recorded by the TCSPC at the relay node. Their time tags are denoted by $T_{Herald}$. Since the optical heralding signal and the photons from the user node propagate along the same fibers, $T_{Herald}$ reflects the arrival time of the single-photon wavepacket from the user node with the impact of length fluctuation of the transmission fibers. Besides, the photons output from the port C7 of the circuit of the relay node are also filtered, detected and recorded by the TCSPC at the relay node. Their time tags are denoted by $T_{Local}$, which indicated the timing reference of the idler photons generated locally in the relay node. By comparing $T_{Herald}$ and $T_{Local}$, $\Delta_{delay}$ can be calculated and used as a feedback signal to adjust the variable optical delay line. By this way, an active feedback control is applied to stabilize the quantum interference in BSM. During the experiment, the active feedback system adjusts the optical delay line every 25 seconds. It is worth noting that two indoor fiber spools are used as the transmission fiber in this experiment. The polarization state of the photons is well preserved when they propagate the optical fibers in the lab environment. Hence, no active feedback method is applied to stabilize their polarization. Details on polarization stability in the experiment is provided in the supplementary information S6.

**Acknowledgements:**

This work was supported by the National Natural Science Foundation of China (92365210, WZ), National Key R&D Program of China (2023YFB2806700, YH), Innovation Program for Quantum Science and Technology (2023ZD0300100, LY), Tsinghua Initiative Scientific Research Program (WZ) and the project of Tsinghua University-Zhuhai Huafa Industrial Share Company Joint Institute for Architecture Optoelectronic Technologies (JIAOT, YH).

**Author contributions:**

W.Z. and D.L. proposed the silicon quantum photonic circuit scheme, were responsible for the chip fabrication and designed the experimental systems. D.L., Z.J., J.L., X.Z., and X.R. carried out the experiment, collected and analyzed the experimental data. H.L. and L.Y. developed and maintained the SNSPDs used in the experiment. W.Z. and D.L. wrote the manuscript. X.F., F.L., K.C. and Y.H. contributed to the revision of the manuscript. W.Z. and Y.H. supervised the whole project. All authors have given approval for the final version of the manuscript.

**Data availability:**

All data needed to evaluate the conclusions in the paper are present in the paper and/or the Supplementary Information. Additional data related to this paper may be requested from the authors.

**Conflict of interest:**

The authors declare no competing interests.

*molecular, and optical physics*, 52, 105-159 (2005).